\documentclass[utf8]{frontiersFPHY} 

\setcitestyle{square} 
\usepackage{url,hyperref,microtype,subcaption}
\usepackage[onehalfspacing]{setspace}

\def\keyFont{\fontsize{8}{11}\helveticabold }
\def\firstAuthorLast{Jin {et~al.}} 
\def\Authors{Kyuho Jin\,$^{1}$ and Unjong Yu\,$^{2,*}$}
\def\Address{$^{1}$Division of Liberal Arts and Sciences, Gwangju Institute of Science and Technology, Gwangju 61005, South Korea \\
$^{2}$Department of Physics and Photon Science, Gwangju Institute of Science and Technology, Gwangju 61005, South Korea }
\def\corrAuthor{Unjong Yu}

\def\corrEmail{uyu@gist.ac.kr}

\begin{document}
\onecolumn
\firstpage{1}

\title[Global Reference and Social Contagion Dynamics]{Reference to Global State and Social Contagion Dynamics} 

\author[\firstAuthorLast ]{\Authors} 
\address{} 
\correspondance{} 

\extraAuth{}

\maketitle

\begin{abstract}

\section{}
The network-based model of social contagion has revolved around information on local interactions; its central focus has been on network topological properties shaping the local interactions and, ultimately, social contagion outcomes. We extend this approach by introducing information on the global state, or global information, into the network-based model and analyzing how it alters social contagion dynamics in six different classes of networks: a two-dimensional square lattice, small-world networks, Erd\H{o}s-R\'{e}nyi networks, regular random networks, Holme-Kim networks, and Barab{\'a}si-Albert networks.
We find that there is an optimal amount of global information that minimizes the time to reach global cascades in highly clustered networks. We also find that global information prolongs the time to hit the tipping point but substantially compresses the time to reach global cascades after then, so that the overall time to reach global cascades can even be shortened under certain conditions. Finally, we show that random links substitute for global information in regulating the social contagion dynamics. 

\tiny
 \keyFont{ \section{Keywords:} Social contagion dynamics, Global information, Global cascades, Complex networks, Agent-based modeling} 
\end{abstract}

\section{Introduction}

Social contagion is commonplace in our daily lives. People decide whether or not to adopt new ideas, beliefs, rumors, norms, business practices, technological standards, or fashion trends~\citep{Morris00,Centola18}. When making this sort of decision, they as cognitive social agents actively refer to or are passively influenced by their neighbors' decisions~\citep{Coleman57}. As such, their social networks play a crucial role in the social contagion process. Through these networks, each decision of the social agents can hold a non-trivial cascading impact on the global contagion process. A research stream has made significant progress along this line by bringing networks as a backbone of the contagion process~\citep{Watts02,Watts07,Centola07,Centola10,Liu20}.

Rooted in the voter model~\citep{Holley75,Watts03,Sood05,Stark08,Yang09} and the majority-vote model~\citep{Oliveira92,Oliveira93,Yu17}, most of the studies on social contagion assume that social agents are short-sighted; their decisions are solely determined by their neighbors' decisions or information on the local adoption ratio (local information, hereafter). In reality, however, social agents also care about and rely at least in part on all other agents' decisions or information on the global adoption ratio (global information, hereafter). Moreover, nowadays, most global information is available and readily accessible as it is increasingly made public almost in real-time in the form of aggregate statistics by intelligence agencies, market research companies, news aggregators, and social media powered by information and communication technology~\citep{Patino12,Poynter10,Cooke08,Domenico20}. The Arab spring is a case in point in that the anti-government protests' spread was instigated and propelled by global information transmitted through social media \citep{Eltantawy11,Wolfsfeld13}. Reliance on global information escalates further to the extent that the pressure for collective behavior primarily operates on a global scale~\citep{Domenico20}. Such global pressure includes legitimating pressure~\citep{Tolbert83}, bandwagon pressure~\citep{Granovetter78,Abrahamson93}, information cascade~\citep{Bikhchandani92}, and network externalities~\citep{Arthur89,Katz85,Shapiro98,Easley10}. 

Nevertheless, limited attention has been paid to how social contagion dynamics will change if knowledge of the global state is incorporated. In this paper, we examine the effects of global information on the social contagion dynamics and attendant outcomes, which also answers the recent call for looking beyond pairwise local interactions in investigating network dynamics \citep{Battiston20,Iacopini19,Ishii,Wang19,Wang21}. For this purpose, we employ the threshold model that enjoys wide currency in the analysis of social contagion. Since the model generally allows for local information only~\citep{Watts07,Holley75,Centola07ajs}, we adapt it to incorporate global information by introducing a relative weight of global to local information. Operationally, we insert the weighted arithmetic mean of the global and local adoption ratio into the threshold model. As a tunable parameter, this relative weight $\gamma$ represents the degree to which a focal system bears the pressure for collective behavior on a global scale.

The effect of global information on the social contagion process can vary according to the topological properties of networks~\citep{Watts03}. Random links (i.e., edges connecting pairs of nodes chosen uniformly at random, be they rewired or newly added) or hubs that dramatically expedite global diffusion may serve as a functional substitute for global information~\citep{Watts98,Barabasi99}. With this possibility in mind, we consider various classes of networks along the dimensions of local clustering, randomness, and scale-freeness. Our results show that the more social agents refer to global information than local information, the slower the social contagion process in most networks. However, in the networks with high local clustering and just a few random links, such as square lattice networks and a subset of small-world networks, reference to global information accelerates social contagion up to a certain point and, after then, decelerates it. This acceleration comes from the breakneck speed of contagion after the tipping point. Nonetheless, an extensive reference to global information increases uncertainties in social contagion outcomes. Finally, random links substitute for global information in the social contagion process.

\section{Model and methods}

We study social contagion based on the threshold model, which considers a network of $N$ agents that interact with one another. Each agent $i$ has his/her own threshold $\theta_i$ and adopts a new option if and only if its adoption ratio is larger than his/her threshold. 
The threshold is often regarded as a constant~\citep{Morris00,Watts02,Choi20PhA}, but we assume that the distribution of the threshold follows the Gaussian function with an average $\mu$ and a standard deviation $\sigma$~\citep{Hackett11}.
The agents with a negative threshold adopt the new option even when the adoption ratio is zero, taking on the role of the instigators.
In contrast, the agents with a threshold larger than one never adopt the new option regardless.
The adoption ratio $a_i$ that an agent $i$ refers to consists of the local and global adoption ratio whose relative weight is determined by a parameter $\gamma$:
\begin{eqnarray}
a_i(t) = (1-\gamma) \frac{d_{i,a}(t-1)}{d_{i}} + \gamma \frac{N_a(t-1)}{N} , \label{adoption_ratio}
\end{eqnarray}
where $d_{i}$ is the degree (the number of nearest neighbors) of the agent $i$. 
It follows that as $\gamma$ increases, the influence of the global adoption ratio vis-\`a-vis the local adoption ratio on his/her adoption decision rises. 
Variables $d_{i,a}(t-1)$ and $N_a(t-1)$ represent the numbers of agents that adopted the new option among the neighbors of the agent $i$ and in the whole network, respectively, at time $t-1$.
At time $t$, all the agents with $a_i(t)>\theta_i$ adopt the new option synchronously. Because an agent's decision at time $t$ depends on his/her neighbors' decisions at time $t-1$, social contagion spreads through the relations between the focal agent and his/her neighbors in a cascading fashion over time. Social contagion ends when no agents are left to adopt the new option. 

Notably, the local threshold model for social contagion centers on the first term on the right-hand side of Eq.~(\ref{adoption_ratio}), i.e., local information, on a specific network~\citep{Centola10,Centola07ajs,Watts02,Watts07}. Therefore, network structure plays a pivotal role in allowing emergent local cascades to spread out and become global cascades. In marked contrast, the global threshold model mostly found in the social science literature focuses on the second term, i.e., global information~\citep{Easley10,Granovetter78,Jackson08}. This model implicitly assumes that any agent can perfectly observe and summarize what others do. A network-based translation of this assumption is that agents are fully connected with one another. As a result, the effect of network structure becomes trivialized. Taken together, Eq.~(\ref{adoption_ratio}) can be interpreted as integrating the two distinct perspectives on social contagion via the tunable parameter, $\gamma$.

Social contagion dynamics are in general analytically intractable in both models, even if exact equations for the dynamics may be derived~\citep{Newman10,Porter16}. This intractability is more pronounced in the local threshold model because the dynamics are compounded by the topological properties of complex networks. For instance, Ref.~\citep{Porter16} analytically derives a mean-field approximation for the dynamics of a local threshold model and numerically shows that its solution represents an S-shaped curve. However, this solution is only applicable to a regular network and in the absence of local clustering and dynamic correlations. Thus, relatively little is known from the local threshold model as yet about social contagion dynamics on diverse networks. All we know for sure is just that networks govern the dynamics by guiding local interactions that buttress the bottom-up process. In this respect, one of this study's contributions is to further enrich our understanding of the local threshold model's dynamics on the basis of numerical analyses.

\begin{figure}
\begin{center}
\includegraphics[width=0.9\columnwidth]{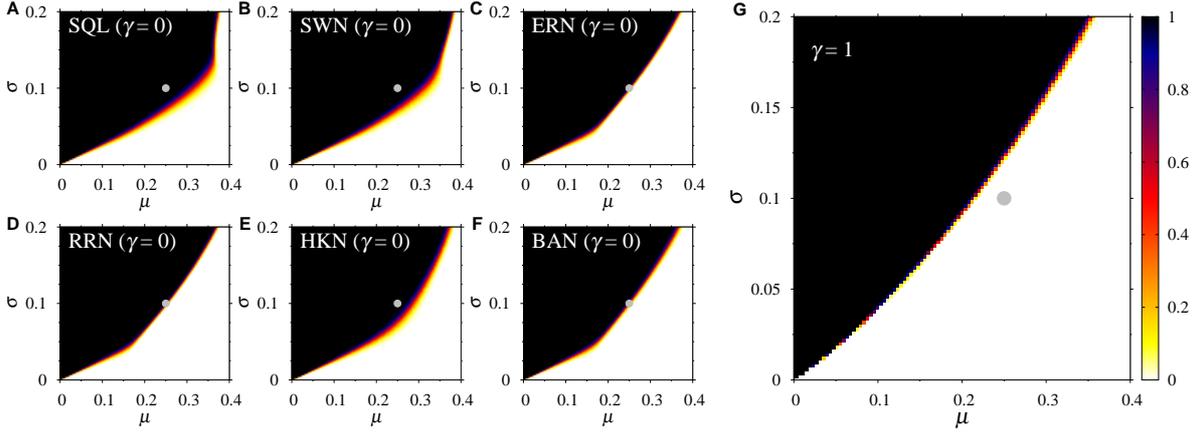}
\end{center}
\caption{The probability of global cascades as a function of an average $\mu$ and a standard deviation $\sigma$ of the Gaussian probability distribution function of the threshold for the six classes of networks with $\gamma=0$ in (\textbf{A})-(\textbf{F}) and that if $\gamma=1$ in (\textbf{G}). (The results do not depend on the network structure when $\gamma=1$.) The rewiring probability $P_{\mathrm{SW}}=0.05$ for SWNs. The gray circle indicates the case of $\mu=0.25$ and $\sigma=0.1$.
}
\label{Fig1}
\end{figure}

Similarly, the equation for the dynamics in the global threshold model is intractable and does not have a closed-form analytic solution although it is represented as a simple recursive formula~\citep{Easley10,Granovetter78,Jackson08}. Nonetheless, it is worth noting that as its dynamics are driven by positive feedback, they often have a tipping (or branch) point from which the system moves away 
upwards or downwards~\citep{Easley10,Jackson08}. Such a bifurcation sends the end state to either a massive social contagion or a tiny unsuccessful one. Therefore, if the dynamics start far below the tipping point as in our case, they are predicted to end up prematurely with a tiny amount of social contagion; in other words, the probability of global cascades approaches zero. This prediction is also consistent with Fig.~\ref{Fig1}(\textsf{\textbf{G}}). Further, if some exogenous shock arises or other forces are introduced to propel the dynamics up to the tipping point, global cascades take place~\citep{Jackson08}, culminating in a massive social contagion.

Now consider Eq.~(\ref{adoption_ratio}) in which these two threshold models are incorporated. Global cascades that were impossible before in the global threshold model can arise because local cascades in the local threshold model are introduced as another driving force for social contagion. The likelihood of and speed with which local cascades propagate and aggregate to become global cascades are governed by $\gamma$ as well as the velocity of local cascades which, in turn, is a function of network topological properties. Given that a large $\gamma$ implies a smaller contribution of local cascades as a kick-starter of the dynamics to social contagion, the likelihood of global cascades is predicted to decrease with $\gamma$. Even so, if local cascades somehow succeed in pushing social contagion up above the tipping point despite the large $\gamma$, the positive feedback gets unleashed, rendering much faster the speed with which local cascades transition into global cascades.

In this study, we consider six different classes of undirected and unweighted networks: a two-dimensional square lattice (SQL), small-world networks (SWNs)~\citep{Watts98}, Erd\H{o}s-R\'{e}nyi networks (ERNs)~\citep{Erdos59}, regular random networks (RRNs)~\citep{Bollobas80}, Holme-Kim networks (HKNs)~\citep{Holme02}, and Barab{\'a}si-Albert networks (BANs)~\citep{Barabasi99}. (See Table~\ref{table_net}.) For SQL, the Moore neighborhood is adopted, so that each agent has eight neighbors. SWNs were obtained by rewiring SQL with the rewiring probability $P_{\mathrm{SW}}$. 
In ERNs, any two agents are linked with the probability $P_{\mathrm{ER}}$, so that the average degree becomes $\langle\Delta\rangle = (N-1) P_{\mathrm{ER}}$. RRNs were generated by the Steger-Wormald algorithm~\citep{Steger99}. 
The two scale-free networks (i.e., HKNs and BANs) were constructed through the growing method \citep{Barabasi99}, such that new agents are introduced one by one with no attrition. Every time a new agent is introduced, a fixed number of links are forged between the new agent and one of the existing agents chosen either by preferential attachment~\citep{Barabasi99} for BANs or by triad formation~\citep{Holme02} for HKNs. Multiple links are excluded. Note that HKNs and BANs have the same degree distribution of $P_d\sim d^{-3}$.
For all the six classes of networks, the number of agents and the average degree are fixed to be $N=10\,000$ and $\langle\Delta\rangle=8$. The effects of varying $N$ and $\langle\Delta\rangle$ are discussed later.
While ERNs, RRNs, and BANs have a vanishing clustering coefficient for large $N$~\citep{Choi20PhA,Holme02,Newman01}, SQL, SWNs, and HKNs are highly clustered. 
The global clustering coefficient $C_g$ and the average local clustering coefficient $C_l$ are defined as
\begin{eqnarray}
&&C_g = \frac{\sum_{i,j,k} A_{ij} A_{jk} A_{ki}}{\sum_{i} d_i (d_i-1)} ,\\
&&C_l = \frac{1}{N} \sum_{i} C_{l,i} ~\mbox{ with } 
C_{l,i} = \left\{ \begin{array}{ll} \frac{\sum_{j,k} A_{ij} A_{jk} A_{ki}}{d_i (d_i-1)} & \mbox{for } d_i>1 \\ 0 & \mbox{for } d_i \leq 1
\end{array}
\right. ,
\end{eqnarray}
where $A$ is the adjacency matrix~\citep{Watts98,Jackson08}. They are $C_g = C_l = 3/7 \approx 0.429$ for SQL, $C_g = 0.365(1)$, $C_l = 0.369(1)$ for SWNs of $P_{\mathrm{SW}}=0.05$, and $C_g = 0.138(3)$, $C_l = 0.536(1)$ for HKNs.
Except for SQL, ten independent networks were generated for each case, and their results were averaged. All the results from the ten different implementations were within the range of the statistical error.

\begin{table}[]
    \centering
    \begin{tabular}{c|cccccc}
    \hline 
                                     & SQL & SWN & ERN & RRN & HKN & BAN \\
    \hline
    Degree distribution          & Regular & Poisson & Poisson & Regular & Scale-free  & Scale-free \\
    Presence of hubs             & No      & No      & No      & No      & Yes         & Yes \\
    Clustering coefficient       & Finite  & Finite  & 0       & 0       & Finite      & 0 \\
    Average shortest path length & Large   & Small   & Small   & Small   & Ultra-small & Ultra-small \\
    Reference                    &  &\citep{Watts98}&\citep{Erdos59}&\citep{Bollobas80}&\citep{Holme02}&\citep{Barabasi99}\\
    \hline
    \end{tabular}
    \caption{Structural properties of the six kinds of networks considered in this work: square lattice (SQL) with the Moore neighborhood, small-world network (SWN), Erd\H{o}s-R\'{e}nyi network (ERN), regular random network (RRN), Holme-Kim network (HKN), and Barab{\'a}si-Albert network (BAN).}
    \label{table_net}
\end{table}

\begin{figure}
\begin{center}
\includegraphics[width=0.9\columnwidth]{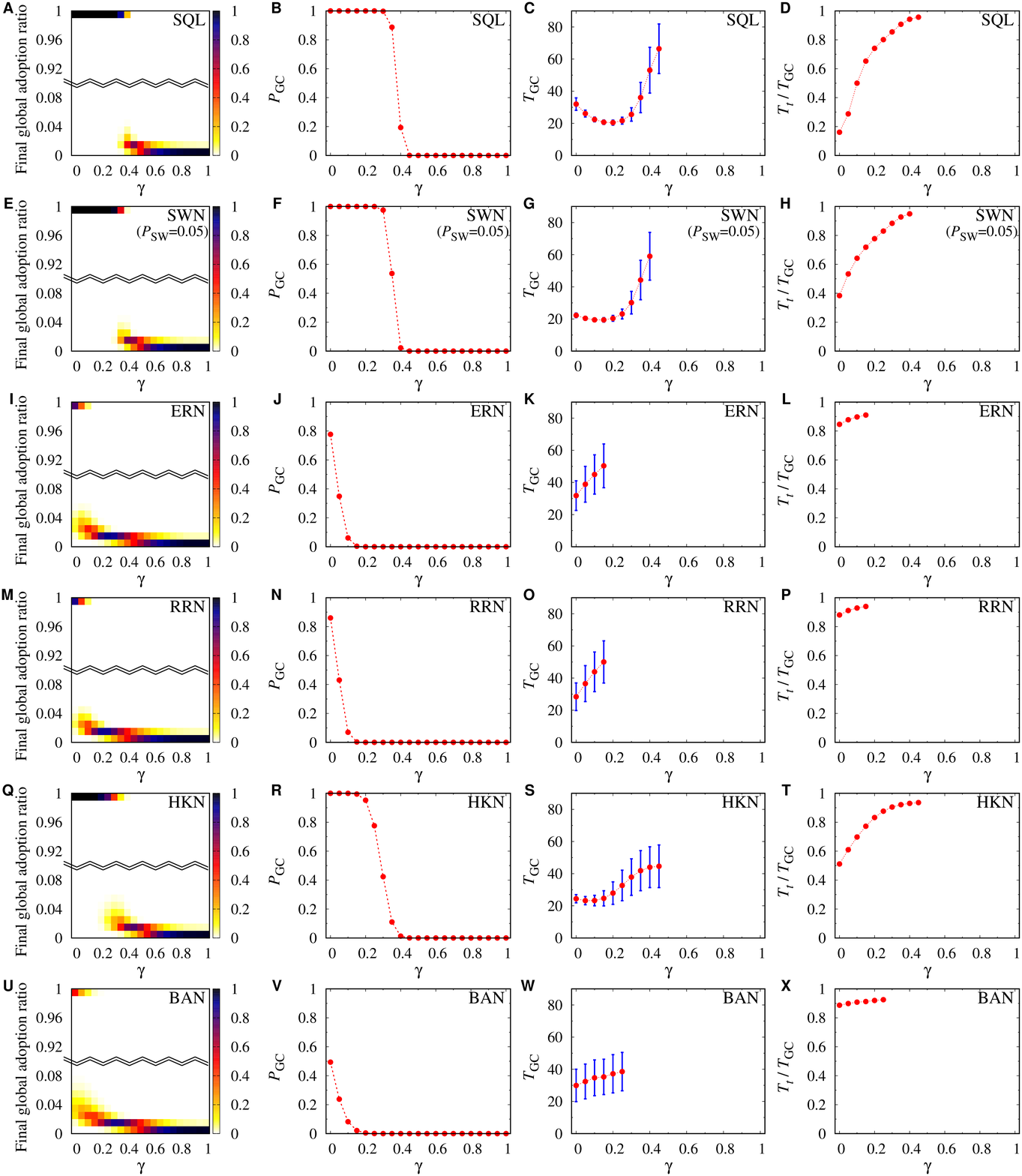}
\end{center}
\caption{The final global adoption ratio, the probability of global cascades ($P_{\mathrm{GC}}$), time to global cascades ($T_{\mathrm{GC}}$), and the ratio of time to tipping points ($T_t$) to time to global cascades ($T_{\mathrm{GC}}$) as a function of $\gamma$ for the six classes of networks.
Global adoption ratio is the number of agents who adopt the new option $N_a$ over the whole population $N$ of the network.
Note that a vertical bar for each symbol of $T_{\mathrm{GC}}$ represents a standard deviation of data, not a statistical error.
As for $P_{\mathrm{GC}}$, $T_{\mathrm{GC}}$, and $T_t/T_{\mathrm{GC}}$, error bars for statistical errors are smaller than the symbol sizes.
}
\label{Fig2}
\end{figure}

\section{Results} 

Figure~\ref{Fig1} exhibits the probability of global cascades in the six different classes of networks as a function of the two parameters, $\mu$ and $\sigma$, of the Gaussian probability distribution function (PDF) of the threshold. The global adoption ratio is not referred to at all ($\gamma=0$) in (\textsf{\textbf{A}})-(\textsf{\textbf{F}}) while being solely referred to ($\gamma=1$) in (\textsf{\textbf{G}}). 
The results are obtained from 1000 independent runs for each parameter set.
Here, global cascades are defined as the final state where the vast majority of the agents adopt the new option.
Operationally, 90\% is used as the cutoff value for determining global cascades. While this cutoff value is admittedly arbitrary, choosing other values like 80\% or 70\% makes no actual difference owing to the sharp phase transition observed in Fig.~\ref{Fig2}.
According to the results, global cascades are more likely in the region with a smaller $\mu$ and a bigger $\sigma$. This is because the number of instigators is larger in the region, given that the number of instigator is on average $(1/2)\{1-\mathrm{erf}[\mu/(\sqrt{2} \sigma)]\}$, where $\mathrm{erf}(z)=(2/\sqrt{\pi}) \int_0^z \exp(-t^2) dt$ is the error function. 
Lines of tipping points for global cascades are observed in all the six networks. That the lines are upward-sloping convex curves means that the probability of global cascades decreases with $\mu$ when the number of instigators is held constant. 
The sharpness of the curves that demarcate the two regions indicates that transitions to global cascades are abrupt.
It is also notable that the region for global cascades is larger in the highly clustered networks (i.e., SQL, SWNs, and HKNs) than in the networks representing low local clustering (i.e., ERNs, RRNs, and BANs). In other words, local clustering promotes global cascades. 
This relationship can be explained by the \emph{detour effect}~\citep{Centola10,Choi20PhA}: each agent in a highly clustered network influences the decision of other agents not just via direct links but also successive indirect links consisting of many short cycles. Therefore, this result suggests that there is a stochastic causal sequence from local clustering through local cascades to global cascades.
Interestingly, these patterns are not influenced by the degree distribution of the networks. For example, RRNs and BANs exhibit similar phase transition patterns even though their degrees follow different distributions\textendash i.e., a homogeneous and a power-law distribution, respectively.

When compared with Fig.~\ref{Fig1}(\textsf{\textbf{A}})-(\textsf{\textbf{F}}), Fig.~\ref{Fig1}(\textsf{\textbf{G}}) demonstrates that full reference to the global adoption ratio non-trivially shrinks the region for global cascades, hinting at the negative effect of global information on global cascades. This is consistent with our prediction offered in the model and methods section. To further investigate the impact of global information on the dynamics of social contagion, we henceforth continue our analysis with the two parameters of the Gaussian PDF fixed at $\mu=0.25$ and $\sigma=0.1$.
While, at this point, global cascades occur with high probability in all the networks unless global information is considered, this point is located in the vicinity of tipping points and thus liable to phase transitions if just a small perturbation is introduced. It is also noteworthy that when only global information is allowed for, the likelihood of global cascades is zero. Taken together, it is inferred that Eq.~(\ref{adoption_ratio}) integrating local and global information via $\gamma$ is well suited to exploring the effect of $\gamma$ on the phase transition when the threshold is based on this PDF.

Figure~\ref{Fig2} shows the results of four social contagion outcomes as a function of $\gamma$, the relative weight of the global to local adoption ratio. The first column displays how the distribution of the final global adoption ratio changes as $\gamma$ increases. Three points are worth mentioning: (i) bifurcation of the social contagion outcome into success and failure of global cascades is observed in all the networks; (ii) the size of global cascades decreases with $\gamma$; and (iii) the networks with high local clustering (i.e., SQL, SWNs, and HKNs) remain more robust to this adverse effect of $\gamma$.
The second column exhibits the probability of global cascades as a function of $\gamma$. As expected before, the probability of global cascades plunges to nearly zero as $\gamma$ increases. Consistent with the prior results, the probability starts to plummet the instant $\gamma$ is introduced in the networks with low local clustering, whereas it remains nearly one up to a certain value of $\gamma$ in the networks with high local clustering.

The third column of Fig.~\ref{Fig2} presents time to global cascades $T_{\mathrm{GC}}$ as a function of $\gamma$. For this, we included only the simulation runs that achieve global cascades. Two aspects are noteworthy. For one, the curves go down and then up in the highly clustered networks (i.e., SQL, SWNs, and HKNs) so as to have an optimal value of $\gamma$ that minimize time to global cascades.
In contrast, the curves monotonically increase in the networks with low local clustering, suggesting that global information just augments $T_{\mathrm{GC}}$.
For another, variance in time to global cascades generally increases with $\gamma$. In our model, the effect of the global adoption ratio substitutes for that of the local adoption ratio in proportion to $\gamma$. In the initial stage of social contagion when just a small cadre of scattered instigators drives local cascades in isolation, the global adoption ratio stays virtually nil. Accordingly, over-reliance on global information in this stage hampers the local adoption process by reducing the set of vulnerable agents ready to change their strategy. Consequently, social contagion propagation appears hard-pressed to experiment with unexplored pathways to global cascades in the reduced vulnerable set, amplifying the variance.

The last column of Fig.~\ref{Fig2} shows the proportion of time to tipping points ($T_t$) to time to global cascades ($T_{\mathrm{GC}}$). In cases where global cascades are achieved, the global adoption ratio non-linearly increases over time, exhibiting an S-shaped curve~\citep{Rogers03,Shimogawa12}. That is, an increment of the global adoption ratio per unit time increases in the former stage and then decreases to zero in the latter stage. We define a tipping point as the global adoption ratio where its second derivative is maximized. So, it is the point at which the acceleration of propagation is the greatest. According to the results in Fig.~\ref{Fig2}, the ratio $T_t / T_{\mathrm{GC}}$ increases with $\gamma$ in all the networks. However, the slope is steeper in the networks with high local clustering. For a large $\gamma$, it takes a considerable time to reach the tipping point, but contrarily very little time to achieve global cascades after the tipping point. To conclude, $\gamma$ decelerates social contagion before the tipping point but explosively accelerates it after then. Thus, our prediction offered in the model and methods section is confirmed again. It is also noteworthy that the influence of $\gamma$ on this divergence in the social contagion speed is more pronounced in SQL, SWNs, and HKNs, the networks with high local clustering.   

\begin{figure}
\begin{center}
\includegraphics[width=0.9\columnwidth]{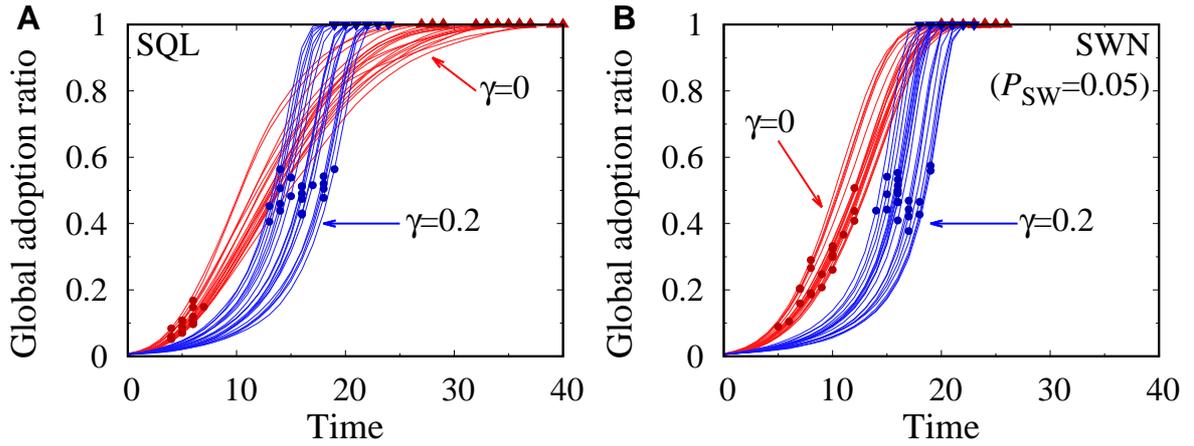}
\end{center}
\caption{Global adoption ratio $N_a(t)/N$ as a function of time for $\gamma=0$ and $\gamma=0.2$ in SQL (\textsf{\textbf{A}}) and in SWNs with $P_{\mathrm{SW}}=0.05$ (\textsf{\textbf{B}}). Twenty social contagion events are plotted for each case. Solid circles and triangles represent the tipping point and global cascades, respectively.
}
\label{Fig3}
\end{figure}

To delve deeper into this diverging social contagion speed or temporal dynamics, we plot the global adoption ratio as a function of time in SQL and SWNs when $\gamma=0$ and $\gamma=0.2$ in Fig.~\ref{Fig3}. 
(According to the third column of Fig.~\ref{Fig2}, the time to reach global cascades is at the minimum around $\gamma=0.2$ in the two networks).
As the figure indicates, the global adoption ratio in the regime of $\gamma=0.2$ falls behind that in the regime of $\gamma=0$ before it hits the tipping point, but, after then, speeds up exponentially and eventually reaches global cascades quicker. 
This could be because, at the optimal point of $\gamma$, most local agents are constrained to remain just vulnerable so as to unleash an explosive chain reaction assisted by positive feedback from $\gamma$ at once if just a small amount of perturbation is added around a tipping point.

\begin{figure}
\begin{center}
\includegraphics[width=0.9\columnwidth]{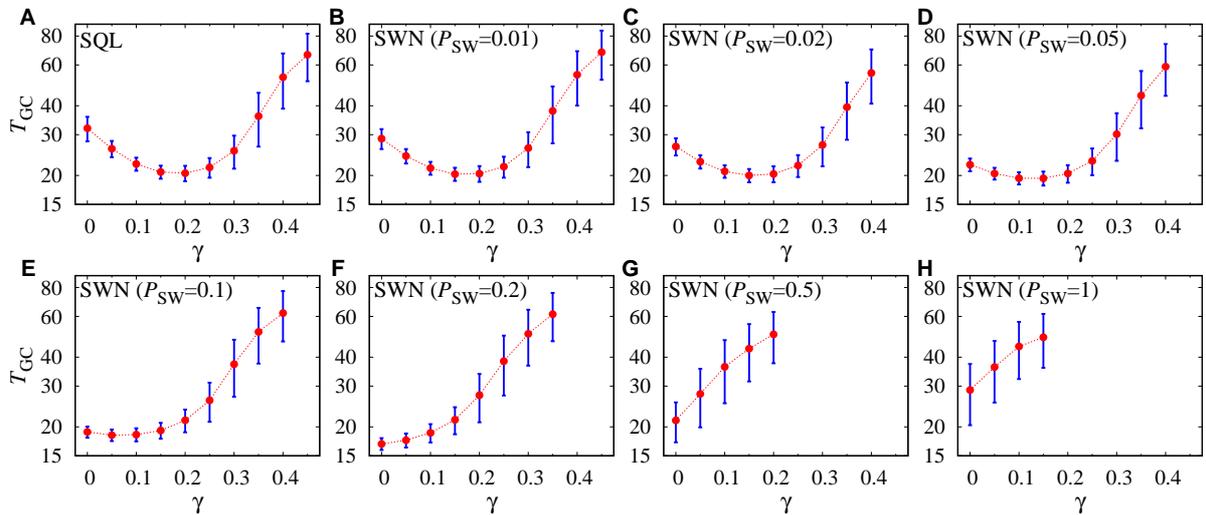}
\end{center}
\caption{Time to global cascades ($T_{\mathrm{GC}}$) as a function of $\gamma$ for the SQL and SWNs with various rewiring probability $P_{\mathrm{SW}}$.
Note that the vertical bars at every symbol represent a standard deviation of $T_{\mathrm{GC}}$, not a statistical error.
Error bars for the statistical error are smaller than the symbol size. 
} 
\label{Fig4}
\end{figure}

Reduction in time to global cascades $T_{\mathrm{GC}}$ and the existence of the optimal $\gamma$ that minimize $T_{\mathrm{GC}}$ are observed only in the highly clustered networks. To understand why, we examined $T_{\mathrm{GC}}$ as a function of $\gamma$ for SQL and SWNs whose rewiring probability $P_{\mathrm{SW}}$ is bracketed from 0.01 to 1, as shown in Fig.~\ref{Fig4}.
In SQL, global information has two countervailing effects: (i) to hinder propagation of adoption cascades by diminishing the effect of local clustering and (ii) to accelerate it by spreading information on the global adoption ratio across the whole network and thereby enabling non-local paths that cannot be uncovered solely by local interactions.
For a small $\gamma$, the latter effect dominates, decreasing $T_{\mathrm{GC}}$. However, as $\gamma$ increases, the former effect gets greater, eventually reducing the likelihood of global cascades and increasing $T_{\mathrm{GC}}$.
In SWNs that are made by adding random links to SQL via rewiring, the latter acceleration effect of global information decreases with the rewiring probability, $P_{\mathrm{SW}}$. Hence, the effect of  $\gamma$ in decreasing $T_{\mathrm{GC}}$ dwindles as $P_{\mathrm{SW}}$ increases, such that $\gamma$ no longer reduces $T_{\mathrm{GC}}$ at any values if $P_{\mathrm{SW}}=0.1$. If $P_{\mathrm{SW}} \geq 0.2$, $T_{\mathrm{GC}}$ monotonically increases regardless of $\gamma$.
These results suggest that random links can substitutes for the effect of global information on global cascades in SWNs by making global information redundant. This mechanism can also explain why the optimal $\gamma$ that minimizes $T_{\mathrm{GC}}$ does not exist in the networks with low clustering (i.e., ERNs, RRNs, and BANs).

In other analyses not reported here, we conducted sensitivity analyses in larger networks whose $N=160\,000$ and in networks whose average degree $\langle\Delta\rangle=6$ and $12$. We found a similar pattern in the larger networks. The suppressing effect of global information on global cascades exists in almost the same region of $\gamma$, but phase transitions occur more abruptly in the larger networks. Consistently, the optimal $\gamma$ that minimizes $T_{\mathrm{GC}}$ is also observed only in the highly clustered networks. We also found that as $\langle\Delta\rangle$ becomes greater, the graphs in Fig.~\ref{Fig2} shift to the left. In other words, increasing $\langle\Delta\rangle$ has the same effect as raising $\gamma$. This is because if $\langle\Delta\rangle$ is increased (i.e., if an agent's ego network is enlarged on average), the agent, given a threshold, requires more neighbors that have already adopted the new option to change his/her mind. In other words, the likelihood as well as effect of local cascades as a kick-starter of social contagion decreases, \emph{Ceteris Paribus}. Because $\gamma$ is a relative effect of global to local information, a decreasing local effect is equivalent to having an increased $\gamma$. Also, for the same reason, the region for global cascades decreases with $\langle\Delta\rangle$.

\section{Discussion}

Social contagion is based on every social agent’s decision. The involved decision is essentially \textit{social}, pressing the social agent to refer to his/her local neighbors as well as the society at large. Decisions of this kind include but are not limited to whether to adopt new technology~\citep{Goolsbee02}, a political view~\citep{Rothschild14}, or organizational structure and strategy~\citep{Tolbert83,Fligstein91,Fligstein93,Haunschild93}, whether to participate in anti-government protests~\citep{Eltantawy11,Granovetter78,Wolfsfeld13} or social movements~\citep{Strodthoff85,Wang12}, and which product to purchase in an ongoing standards war observed in the market for operating systems (e.g., iOS vs. Android), messaging apps (e.g., WhatsApp vs. Snapchat), and future mobility choices (e.g., electric cars vs. hydrogen cars)~\citep{Katz86,Shapiro99}. Notably, all these decisions influence not just each agent’s social and economic (microscopic) well-being but society’s (macroscopic) welfare as a whole. Therefore, it is vital to provide social agents and social system designers like government officials, policymakers, or social activists with more precise knowledge of how social contagion unfolds over time. Nonetheless, the extant studies are less successful in doing so because they focus exclusively on one information by neglect of the other.

In this spirit, we attempt to integrate local and global information in analyzing social contagion dynamics. Specifically, revising the majority-vote model that considers only local interactions, we have investigated the role of global information in the social contagion dynamics on various classes of networks. The results show that global information accelerates the social contagion process if two conditions are met at once: (i) the networks for social contagion are highly clustered and (ii) only a small amount of global information is introduced. Social contagion is complex and therefore necessitates information reinforcements from multiple local neighbors. In this respect, local clustering is conducive to social contagion as it maximizes local information reinforcements via the detour effect. However, local clustering can reduce the social contagion speed by forcing it to take on the locally connected routes. Our results demonstrate that global information counteracts this deceleration effect of local clustering and even accelerates the social contagion speed by providing alternative non-local shorter paths. This implies that global information works as a functional substitute for random links even if their underlying mechanisms differ. Nevertheless, excessive global information weakens the role of local interactions--the engine for complex social contagion--and therefore decelerates the social contagion speed. It follows that there is an optimal amount of global information to minimize the time to reach global cascades in the highly clustered networks.

These results also hold many real-world implications for individual agents and social architects. For instance, consumers are advised to consider buying a product in an ongoing standards war if the product appears to keep inching its way across the densely connected local networks; corporate managers who want to avoid a newly launched product’s premature death are encouraged to keep investing resources in gaining the product’s acceptance in the local networks for a longer time than previously thought if the target customers care about the global adoption ratio; electric vehicle makers should try their best to ramp up charging station infrastructure not to benefit from the network externalities but because it reduces the negative effect of the global adoption ratio on global cascades in the initial period; government agencies for environmental protection are recommended to remain steadfast in offering incentives such as subsidies and tax credits for eco-friendly vehicles to ensure that local cascades keep spreading until the tipping point for global cascades is reached; social movement activists may have to focus more attention on a grassroots movement by cultivating and tightening their grip on the local community and networks as they allow local cascades to leapfrog into global cascades. As such, our results have significant relevance to the real world and proffer practical advice and recommendations.

\section{Limitations and Directions for Future Research}
Our study is based on numerical analyses of the simulation model. While this approach enjoys numerous virtues particularly when it is difficult to derive analytical solutions, future studies should continue efforts on analytic approaches that are deemed to offer more insights and generalizability~\citep{Newman10}. Several studies have made progress in this direction~\citep{Gleeson08,Gleeson11,Gleeson13,Sullivan15}. In addition, while we conducted our simulations on the six generic classes of networks with a view to approximating a controlled experiment design, the networks are not real-world networks. Future studies may pursue new findings by re-conducting our analysis on real-world networks. Besides, our results depend on one probability distribution function of the threshold—the Gaussian function. While the Gaussian function is widely used in the social contagion literature~\citep{Moore91,Rogers03}, social contagion may manifest non-trivially different dynamics if the threshold follows a different probability distribution. Further, our model implicitly assumes that the global adoption ratio is precise and correct. However, this may not always be true; the information could be distorted inadvertently or deliberately by information providers. Incorporating this possibility into our model may not just produce practically interesting results but also improve the model’s predictive validity. Relatedly, it is also often the case that global information is not freely available but costly to obtain while the agent’s ability to obtain it is not uniformly distributed. Probing into this possibility holds promise for enriching our knowledge. Also, whereas our study is based on unweighted networks, most real-world networks are weighted ones~\citep{Bellingeri20}. It may non-trivially shift our findings and thereby help identify a new research direction to consider weighted networks. Future studies would benefit from entertaining this possibility. Finally, our analysis is basically predicated on the single-layer network, which could be viewed as a simplifying projection of various networks in operation and diverging interactions therein onto the network of the focal layer~\citep{Boccaletti14,Wang15,Wang19,Wang21}. It could be that social contagion propagating on the real-world network is preceded and driven by that on the networks of online social media or other types of networks. Then, an investigation into the root cause of social contagion would call for an alternative theoretical lens. Future studies can benefit from examining this possibility and yield new insights along the way.

\section{Conclusion}
Most studies on social contagion have examined the role of local information to the exclusion of global information or the other way around. However, simultaneous consideration of both information gives rise to non-trivially different social contagion dynamics. For instance, the two different kinds of information exhibit complementarity, albeit conditional, in facilitating the social contagion process. Also, global information extends the time to hit the tipping points and, at the same time, substantially compresses the time to reach global cascades afterward. Further, an excessive reference to global information adds uncertainties to the social contagion process. Therefore, our results suggest that social contagion may be more complex than expected if global information assumes a certain role in the process. We believe that this consideration is opportune, in view of the rapid advancement of information and communication technology that expedites the generation and distribution of global information across the world. It is hoped that the introduction of global information provides an impetus for further analyses on social contagion that better reflect the reality in a constant state of flux. 

\section*{Conflict of Interest Statement}

The authors declare that the research was conducted in the absence of any commercial or financial relationships that could be construed as a potential conflict of interest.

\section*{Author Contributions}

KJ conceived and designed the study and wrote the original draft. UY developed the code and performed the simulations. Both authors interpreted the results and reviewed and modified the final draft.

\section*{Funding}

This work was supported by GIST Research Institute (GRI) grant funded by the GIST in 2021.

\bibliographystyle{frontiersinHLTH&FPHY} 
\bibliography{ref_global}

\end{document}